# 4MOST Survey Strategy Plan

Guillaume Guiglion[1]
Chiara Battistini[2]
Cameron P. M. Bell[1]
Thomas Bensby[3]
Thomas Boller[4]
Cristina Chiappini[1]
Johan Comparat[4]
Norbert Christlieb[2]
Ross Church[3]
Maria-Rosa L. Cioni[1]
Luke Davies[5]
Tom Dwelly[4]
Roelof S. de Jong[1]
Sofia Feltzing[3]
Alain Gueguen[4]
Louise Howes[3]
Mike Irwin[6]
Iryna Kushniruk[3]
Man I Lam[1]
Jochen Liske[7]
Richard McMahon[6]
Andrea Merloni[4]
Peder Norberg[8]
Aaron S. G. Robotham[5]
Olivier Schnurr[1,9]
Jenny G. Sorce[10,1]
Else Starkenburg[1]
Jesper Storm[1]
Elizabeth Swann[11]
Elmo Tempel[12,1]
Wing-Fai Thi[4]
C. Clare Worley[6]
C. Jakob Walcher[1]
and The 4MOST Collaboration

[1] Leibniz-Institut für Astrophysik Potsdam (AIP), Germany
[2] Zentrum für Astronomie der Universität Heidelberg/Landessternwarte, Germany
[3] Lund Observatory, Lund University, Sweden
[4] Max-Planck-Institut für extraterrestrische Physik, Garching, Germany
[5] International Centre for Radio Astronomy Research/University of Western Australia, Perth, Australia
[6] Institute of Astronomy, University of Cambridge, UK
[7] Hamburger Sternwarte, Universität Hamburg, Germany
[8] Department of Physics, Durham University, UK
[9] Cherenkov Telescope Array Observatory, Bologna, Italy
[10] Centre de Recherche Astrophysique de Lyon, France
[11] Institute of Cosmology and Gravitation, University of Portsmouth, UK
[12] Tartu Observatory, University of Tartu, Estonia

The current status of and motivation for the 4MOST survey strategy, as developed by the Consortium science team, are presented here. Key elements of the strategy are described, such as sky coverage, number of visits and total exposure times in different parts of the sky, and how to deal with different observing conditions. The task of organising the strategy is not simple, with many different surveys that have vastly different target brightnesses and densities, sample completeness levels, and signal-to-noise requirements. We introduce here a number of concepts that we will use to ensure all surveys are optimised. Astronomers who are planning to submit a Participating Survey proposal are strongly encouraged to read this article and any relevant 4MOST Survey articles in this issue of The Messenger such that they can optimally complement and benefit from the planned surveys of the 4MOST Consortium.

4MOST is a new wide-field spectroscopic survey facility to be mounted on the 4-metre VISTA telescope. Observations are expected to start in 2022 after which 4MOST will be running multiple survey periods, each of a five-year duration. More details of the 4MOST instrument and proposal submission process can be found in the 4MOST overview paper (De Jong et al., p. 3). For more details of scientific operations, we refer the reader to the 4MOST Scientific Operations paper (Walcher et al., p. 12).

The multiplex of 4MOST is so large that few science cases have sufficiently high target densities to fill all the fibres in a 4MOST field of view on their own. There are many important science cases that need only a few targets in each field of view but have targets spread over the entire sky. To efficiently fill all the fibres and to make low-target-density surveys possible, it was realised early on that 4MOST would benefit from an operations scheme in which most 4MOST science programmes are merged into one survey and observed simultaneously. The use of the 4MOST Guaranteed Time Observations was developed around this concept and constitutes the Consortium Survey Plan presented here. The ESO community is invited to participate in this joint strategy, giving such Participating Surveys the added benefit that the exposure time of targets in common with other surveys will be shared proportionately.

The current survey plan is subject to change due to several factors. Most importantly, the new Community Surveys that will be added to the overall 4MOST survey programme following the proposal process led by ESO will most likely introduce new strategy requirements. Changes can also be expected when further knowledge is obtained regarding the sensitivity of the instrument and its overheads performance as it is further developed, built and tested. Therefore some of the information in what follows is preliminary and subject to change. Up-to-date information on 4MOST, its performance and Survey Strategy will be continuously posted on the 4MOST webpage[1].

## Basic concepts and default strategy

All 4MOST observing programmes are carried out by surveys, each of which can consist of several sub-surveys. Running 4MOST efficiently with many surveys in parallel demands an observing plan that can accommodate targets requiring very different exposure times. Some of the brightest stars may reach their required signal-to-noise (S/N) ratios in five minutes using the Low Resolution Spectrograph (LRS), whereas faint extragalactic targets or faint stars observed with the High Resolution Spectrograph (HRS) could require two hours of total exposure time. The operations scheme must also be able to adapt to different observing conditions (for example, sky brightness and seeing).

To accommodate the different exposure times, the total exposure time in an area on the sky is broken up into individual exposures. For each exposure, the fibres are repositioned in a new configuration such that targets that can be finished in one exposure receive a fibre only once, while other targets receive a fibre multiple times in the different fibre configurations until they reach the required S/N. To save the overheads of repositioning the telescope and acquiring guide stars, several fibre configuration exposures can be grouped into one Observation Block (OB) that is then observed in one telescope visit at that pointing. While we expect that exposure times will be typically of order 20 minutes, shorter times will be used in areas with many bright sources at the cost of some extra overheads. The maximum individual exposure time for a configuration is set at 30 minutes, because changing differential refraction across the field would cause fibres to drift if exposed much longer.

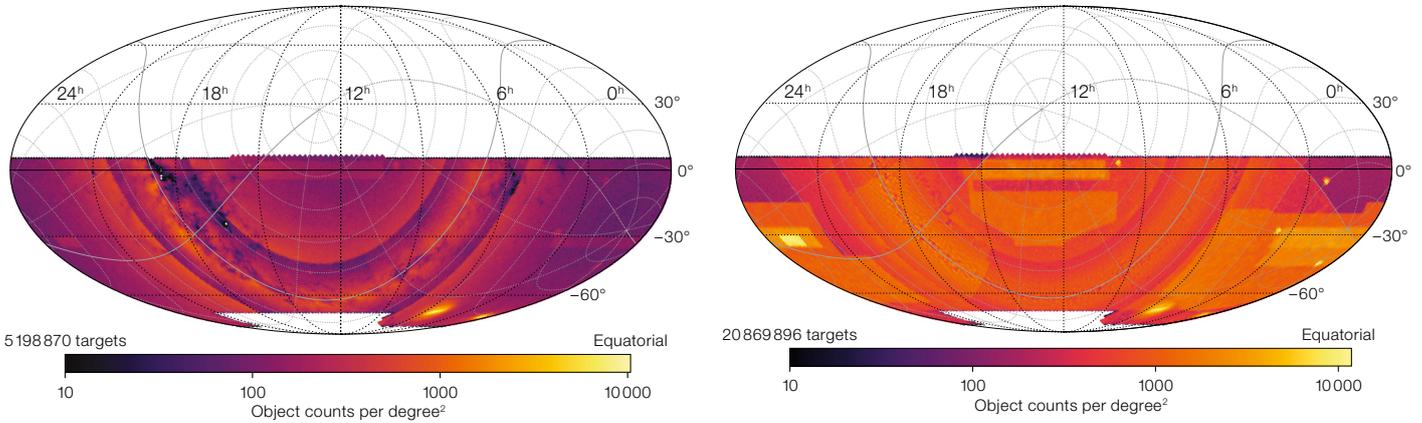

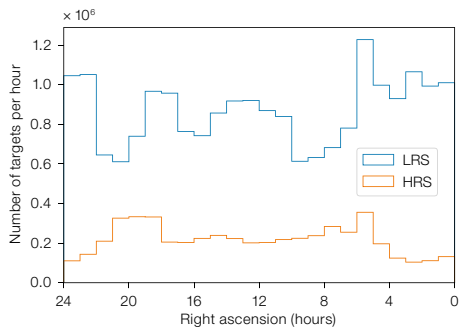

Figure 1. The summed target densities of Consortium Surveys for the HRS (upper left) and the LRS (upper right) sampled to the requested completeness, as a function of RA and dec. The area covered is representative the current survey plan. The lower panel presents histograms of the targets as a function of RA.

The fibre usage efficiency of a grid-based fibre positioner such as that used by 4MOST is increased when the target density is significantly higher than the fibre density in the field of view. Therefore, science cases that can provide many more interchangeable targets than needed to fulfil the science case increase the efficiency of the joint survey programme.

To get a feeling for what is possible with 4MOST, we can assume a baseline strategy of visiting a large fraction of the southern sky twice, with each visit having three fibre configurations exposed for 20 minutes each. Given that Paranal provides about 300 useable nights per year with on average duration of 9 hours, along with 4MOST's field of view of 4.2 square degrees, we can expect to cover about 21 000 square degrees in a five-year survey with two hours total exposure time, assuming about 75% effective open shutter time. Such a basic strategy would allow one to cover the entire sky in the declination range –70 < dec < 5 degrees. This particular preferred declination range was chosen for two reasons. Firstly, one needs to avoid having to cover too much sky in the north where observing time is limited by the regular occurrence of strong northerly winds at Paranal. Secondly, the 4MOST Atmospheric Dispersion Corrector only functions up to a 55-degree zenith distance and hence observations at airmasses larger than 1.75 will see significant sensitivity losses at the ends of the spectral range. Therefore, large numbers of particularly long observations at dec < –70 degrees should be avoided.

To cover special areas on the sky with exposures longer than two hours or outside this fiducial declination range therefore requires giving up coverage and/or total exposure time within certain areas of this sky declination range. A description of the relative coverage of different regions by the 4MOST Consortium Surveys is provided below.

Observing conditions

Because many targets from different surveys are observed simultaneously, proposers cannot request specific observing conditions (seeing, Moon) on a per target level. In order to avoid having too wide a range in target brightness in one area of the sky and to simplify scheduling, the Consortium has identified the disc plane of the Milky Way as the region that will predominantly be observed during bright time, with the rest of the sky devoted mostly to grey and dark time. This means that targets with a fibre luminosity fainter than that of the bright sky will be hard to schedule at low Galactic latitude, with the possible exception of some areas near the bulge where some dark/grey time will be used. The algorithm to assign targets and sky areas to dark/grey/bright observing time is still being improved and hence current diagrams are based on a simplified approach and are only indicative of the final strategy.

Observers will not be able to specify seeing conditions for their observations. When necessary, longer exposure times will be used for a field to match the observing conditions. On a larger scale, we expect to tune the scheduling algorithm such that areas with many background-limited, point-like sources or regions that are hard to complete (for example, high-airmass or high-right-ascension [RA] pressure regions) will automatically be assigned better seeing conditions.

### Encoding the scientific drivers of the survey strategy

In order to have an automatic routine that can develop an observing strategy that is both efficient and reproducible at least in a statistical sense, surveys need to provide a number of other parameters alongside their target catalogues, such that target selections and observing schedules can be optimised. These parameters are described in this section.

In this framework, the Spectral Success Criteria (SSC) were defined at the target level and are used to set the initial exposure time requirement. The SSC prescribe the S/N requirements in specified wavelength regions of the target spectrum needed for robust scientific output. As an example, to achieve a sufficient precision in the stellar parameters, the median S/N ratios in the continuum over given wavelength ranges have to reach at least a certain value. Such criteria are provided by all individual 4MOST Surveys for all of their targets and depend on their individual scientific goals. 4MOST operations are also expected to include a feedback loop on already observed targets. Surveys must therefore also provide "stop observing" criteria for all targets. These are evaluated after the first exposure(s) of a target have been taken in one OB and may be used to make decisions if subsequent observations in this region of the sky are planned. Examples of stop criteria are: a minimum S/N has been reached (which can be lower than the original request); a redshift has been obtained; or a maximum exposure time has been exceeded.

The Small Scale Merit (SSM) is used by the target scheduling tool that assigns fibres to targets to quantify the success of observations in a small area of the sky. The SSM defines the completeness requirements of a (sub-)survey on a scale of one fibre configuration, i.e., an area covering a 4MOST field of view. Indeed, the different surveys require different completeness in a given local area, with differing numbers of targets. The SSM is then a way to quantify the increment of scientific knowledge we acquire when observing an additional target in one survey versus another. This allows surveys to provide many more targets than needed for their science case and, by specifying that only a fraction of targets need to be completed, helps improve the fibre usage efficiency. Using the total observing time assigned to a region and the target exposure times calculated with the SSCs, an algorithm is used to assign targets to fibres in a field of view in a probabilistic way such that the desired completeness is reached for each survey, while avoiding unwanted biases in brightness or crowding, for example. The use of probabilities throughout selection and operation decisions will greatly simplify re-creating the selection functions that one needs to make statistical inferences on the intrinsic abundances of different types of targets.

In order to coordinate the science goals over a large area of the sky, the Consortium uses the Large Scale Merit (LSM). The LSM concerns the entire observable sky and is defined by a HEALPix[a,2] map of scientific priorities as a function of right ascension and declination. The LSM maps are needed to ensure that observations concentrate on areas of higher scientific interest. As an example, surveys can use the LSM to reduce the priority of regions with high levels of extinction.

In order to obtain an overall measure of 4MOST survey success a figure of merit (FoM) is defined by each survey. This metric ranges from 0.0 to 1.0, and can be a function of the successfully observed targets, areas completed with sufficient number of targets, and completeness of individual sub-surveys with special targets. The FoM is defined such that it reaches 0.5 once a survey has met its requirements (the minimum set of observations to accomplish the core science case) and it reaches 1.0 once it has met all its goals. The goal of the 4MOST observation scheduling software is of course to maximise this FoM for all surveys, without penalising any of the surveys. The choice and implementation of these LSM and SSM concepts contribute to the final shape of the 4MOST sky.

The accompanying white papers from the different 4MOST Surveys in this issue of The Messenger provide examples of SSC and FoM usage.

### Survey strategy simulations

To check the feasibility of the 4MOST strategy, the project developed the 4MOST Facility Simulator (4FS), a software tool used to simulate the progress of the five-year 4MOST Survey. The 4FS is used to plan, optimise and verify many aspects of operations planning. The 4FS contains a detailed parameterised model of the 4MOST Facility, including representations of the instrument focal plane, the various constraints/limitations, and a statistical model of the operating environment (for example, the long-term atmospheric/environmental conditions at Paranal, maintenance procedures, etc.). Where possible, the 4FS uses prototype versions of the various algorithms that will be used to operate the real 4MOST Survey, for example, fibre-to-target assignment, survey strategy, survey scheduling, and progress balancing/feedback algorithms. The 4FS serves as the primary test-bed for development and optimisation of these algorithms and their optimal control parameters.

The simulator provides the means to inspect the simulated 4MOST performance on many levels, from individual target classes in sub-surveys, to survey completeness levels, fibre efficiency, and tension between different survey requirements in different parts of the sky and with time. Here we present only a few overview plots of the current Consortium Survey plan, which will be further optimised once combined with the Community Surveys.

First, we present in Figure 1 the requested density of targets as a function of RA and dec as well as a histogram of requested targets as a function of RA. These target densities have been built from preliminary target catalogues submitted by all Consortium Surveys, Galactic and extragalactic jointly, and limited to areas that are in the current survey plan. Community Surveys that match these target distributions will be easiest to schedule.

Figure 2 shows the exposure time distribution over the sky in a recent realisation of a five-year survey with 4FS. The fiducial survey strategy of 4MOST foresees that every field will be observed twice with Observation Blocks containing three fibre configurations with exposure times of about 20 minutes each. However, as mentioned before, the survey teams have identified the need for special survey areas that have higher requested target densities and/or have fainter targets and hence require more exposure time. These regions and their approximate total exposure times are listed in Table 1. To make this possible, the current strategy avoids, or strongly reduces, the amount of exposure time in high extinction areas in the Milky Way disc. In addition, some areas in the –10 < dec < 5 degree range have reduced exposure times as there are no targets available in these areas from the X-ray eROSITA Surveys (S5 Clusters and S6 active galactic nuclei [AGN]).

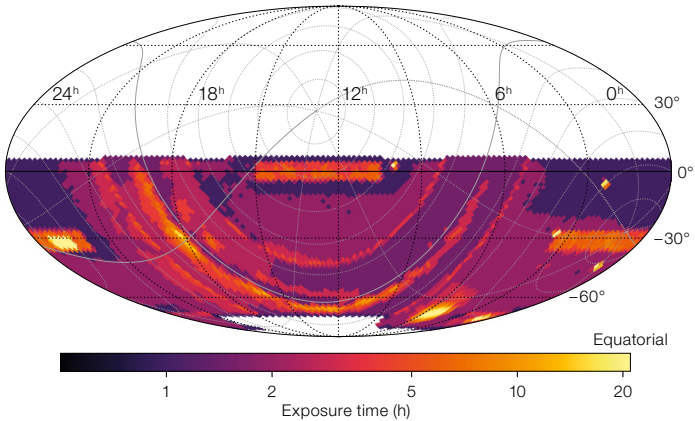

Figure 2. The total exposure time in various areas of the sky in a recent simulation with the 4MOST Facility Simulator.

| Location | Area (square degrees) | Average $t_{exp}$ (hours) |
|---|---|---|
| Bulge and Inner Galaxy | 500 | 4–6 |
| Magellanic Clouds | 200–300 | 2–10 |
| WAVES-Wide | 1300 | 3–4 |
| WAVES-Deep | 50 | 7 |
| LSST Deep Drilling Fields | 4 × 4.2 | 4 × 60 |
| South Ecliptic Pole area | 300 | 4 |

Table 1. Planned special areas in the survey plan with their approximate total area and typical total exposure time per pointing.

The colour code of the map of Figure 2 indicates the total exposure time allocated to each part of the sky. Typically, the stellar (i.e., Galactic) surveys have brighter targets while the extragalactic ones have many targets that cannot be observed when the Moon is present. Therefore, most of the bright time is dedicated to the Milky Way disc regions. However, some dark/grey time will be allocated to mostly stellar fields as well as to enable deeper observations, for instance in the Milky Way bulge and the Magellanic Clouds. Conversely, when the Milky Way disc is not visible bright time will be used at higher Galactic latitude to observe bright targets there.

Table 2 lists the approximate total numbers of targets that are expected to be observed by the individual surveys and that were shown to be feasible in a recent 4FS simulation based on the preliminary target catalogues. While these numbers are close to the goals of each survey, further tuning and balancing between the surveys will lead to subsequent modification of the final statistics.

In running the 4FS it is assumed that the survey strategy remains the same over the entire five-year period. In practice this may not be fully tenable and a feedback loop foresees that adjustments may be made on about a yearly timescale to ensure that all surveys progress on a common scale. However, changes, especially also on the target input catalogues, are expected to be kept to a minimum to ensure that the calculation of selection functions are tractable.

A number of other key elements pertinent to 4MOST Surveys that cannot be demonstrated with this high-level overview are described in the next sections.

## Cadence and time variable sources

As 4MOST is a multi-object survey facility, it cannot deliver timed observations for individual targets. However, using minimal constraints on the scheduling of the observations, variable sources and transients are part of the science cases for some of the 4MOST Surveys. In order to deliver reliable scientific output for these targets, the 4MOST Survey needs to adopt a cadence in its observations. The executed time sequence may be irregular and there can be no guarantee that all repeats will be performed at the scheduled time. The schedule of re-observation of targets is independent of the final total S/N ratio. However, in many cases there is a requirement on the minimum S/N ratio to be reached for a given visit.

Two Consortium Surveys require particular cadences: 1) TiDES, which aims at following up variable AGN and LSST transients, covering LSST Deep Drilling Fields; and 2) the Magellanic Clouds Survey (1001MC) for the pulsating variables stars. The Galactic Surveys can also benefit from the observing cadence as the plan is to observe the whole 4MOST sky at least twice, and, by assuring that there is at least one year between revisits rather than a few months, a larger fraction of binary stars can be identified by their variable radial velocities.

The planned deep fields are natural places to observe variable objects requiring repeat visits. While it will not be possible to perform timed observations, it is expected that a minimum and possibly a maximum duration between repeats can be specified to drive the schedule. The goal is for instance to observe the four declared LSST Deep Drilling Fields about every two weeks plus or minus a few days in order to perform reverberation mapping of AGN. If LSST decides to define another Deep Drilling Field in the South Ecliptic Pole area, this field may replace one of the currently defined four fields. Also repeat fields in the Milky Way bulge area will be used to study variable objects.

Another class of time-variable sources are transients. It will be possible to add a small number of transients, such as recently discovered supernovae, to a survey catalogue over the course of the five-year survey period. Because OBs that drive the observations on the telescope are re-created in intervals of three days, this is also the lead time for adding transient objects. Data from the LSST survey will provide several live transients with durations of several weeks that are visible within one 4MOST field of view at any pointing on the sky. These transients are bright enough that, by adding these targets to the 4MOST target catalogue with high priority, spectra will be obtained by targeting several transients in each observation, following the nominal survey plan of 4MOST.

| Consortium Survey | | Brightness range (magnitudes) | Targets (millions) |
|---|---|---|---|
| S1 | Milky Way Halo LR | $15.0 \leq G \leq 20.0$ | 1.5 |
| S2 | Milky Way Halo HR | $12.0 \leq G \leq 17.0$ | 1.5 |
| S3 | Milky Way Disc and Bulge LR (4MIDABLE-LR) | $14.0 \leq G \leq 19.0$ | 10.0 |
| S4 | Milky Way Disc and Bulge HR (4MIDABLE-HR) | $10.0 \leq G \leq 15.5$ | 2.5 |
| S5 | Galaxy Clusters | $18.0 \leq r \leq 22.0$ | 1.7 |
| S6 | AGN | $18.0 \leq r \leq 22.8$ | 1.0 |
| S7 | Galaxy Evolution (WAVES) | $18.0 \leq r \leq 22.5$ | 1.6 |
| S8 | Cosmology Redshift Survey | $20.0 \leq r \leq 23.9$ | 8.0 |
| S9 | Magellanic Clouds (1001MC) | $10.5 \leq G \leq 19.5$ | 0.5 |
| S10 | Transients (TiDES) | $18.0 \leq r \leq 22.5$ | 0.3 |
| Total | | | > 28 |

Table 2. The minimal number and typical magnitude range of targets that each Consortium Survey expects to observe in the first five-year survey of 4MOST.

Finally, several surveys wish to optimise the sky coverage such that a contiguous area is covered, which has implications for the cadence. The optimisation process for preferentially covering large ($\geq 500$ square degrees), contiguous areas of the sky is still under development.

### Calibration

An important ingredient of the 4MOST survey strategy is the calibration plan. The instrument and data calibration will follow standard procedures for multi-object spectroscopic instruments. The moving spine principle of the positioner will cause variability in the throughput of individual fibres for each science exposure. Therefore the instrument has the capability to take one additional calibration (a fibre flat) per science exposure during the night if needed. Relative flux calibration as a function of wavelength will be performed with the help of white dwarf stars and Gaia spectrophotometry, which will be available at the time 4MOST starts observing. A precision of the order of 10% in the continuum slope is expected.

For sky subtraction a fixed percentage of fibres will be allocated to empty sky regions. To monitor the quality and correctness of the data reduction process, radial velocity, flux, and telluric standard stars will be observed. Commissioning will be used to test the data reduction and analysis pipelines on real data.

### Supplementary targets

Normal completeness requirements of surveys that are used for statistical inferences result in inefficiencies in assigning all fibres to targets once most targets have been completed. In order to increase the 4MOST scientific outcome, the science team will add targets to fill these unscheduled fibres. Supplementary targets are targets that come with no completeness requirements and no guarantee that any one in particular will be observed. These extra targets are only added to avoid empty fibres. Therefore the observed number of such targets in any region will depend on the availability of main survey targets and observing time in a given area. Community proposals for supplementary targets will also be accepted.

### Poor observing conditions programme

When observing conditions are too poor to carry out the normal survey programme, 4MOST will switch to a dedicated poor conditions programme. Poor conditions are, for instance: twilight, full moon without a visible Milky Way, seeing full width half maximum (FWHM) > 1.5 arcseconds, and cirrus. The optimal boundaries for each of these constraints that are required to switch to this special programme have yet to be determined. This programme consists of all stars in Gaia Data Release 3 (DR3) with dec < +30 degrees and in the brightness range $7.5 < G < 11$ magnitudes for HRS, and $11 < G < 12.5$ magnitudes for LRS. This will ensure 4MOST spectroscopy for all stars that form the core samples of the TESS and PLATO space missions. These missions will provide key asteroseismology information that can be used not only to improve the precision on the derived stellar parameters and chemical abundances, but also to compute masses and ages. This makes this sample an ideal calibration and training set for the entire 4MOST Survey. It is expected that almost all of these stars can be observed in less than five years, after which a fraction of them will be regularly repeated with a cadence that still has to be determined.

### Prospects

The 4MOST strategy described in this document is not yet frozen and will need further optimisation. Regular updates will be made, especially when accommodating accepted community proposals. Also, a better understanding of instrument performances will impact the final strategy, as well as further advances in the fields of the presented science cases. As the 4MOST project progresses towards first light more features will be added to the 4MOST Facility Simulator to increase the fidelity of real observations and to encode more science drivers from the surveys. The latest released survey strategy plan can be found via a webpage[3] that will be regularly updated.

### Links

[1] The 4MOST webpage: www.4most.eu
[2] HEALPix webpage: https://healpix.jpl.nasa.gov/
[3] 4MOST Survey Simulations webpage: https://www.4most.eu/cms/operation/simulations/

### Notes

[a] HEALPix is an acronym for Hierarchical Equal Area isoLatitude Pixelization of a sphere, a pixelisation procedure that produces a subdivision of a spherical surface in which each pixel covers the same surface area as every other pixel.